\documentclass[twocolumn]{aastex62}
\usepackage{amsmath}
\usepackage{longtable}
\received{today}
\submitjournal{ApJ}

\shorttitle{Chemical Signatures of the FU~Ori Outbursts}
\shortauthors{Molyarova et al.}

\begin{document}

\title{Chemical Signatures of the FU~Ori Outbursts}

\correspondingauthor{Tamara Molyarova}
\email{molyarova@inasan.ru}

\author{Tamara Molyarova}
\affil{Institute of Astronomy, Russian Academy of Sciences, 48 Pyatnitskaya St., Moscow, 119017, Russia}

\author{Vitaly~Akimkin}
\affiliation{Institute of Astronomy, Russian Academy of Sciences, 48 Pyatnitskaya St., Moscow, 119017, Russia}

\author{Dmitry~Semenov}
\affiliation{Chemistry Department, Ludwig Maximilian University, Butenandtstr. 5-13, D-81377 Munich, Germany}
\affiliation{Max Planck Institute for Astronomy, K{\"o}nigstuhl 17, 69117 Heidelberg, Germany}

\author{P\'{e}ter~\'{A}brah\'{a}m}
\affiliation{Konkoly Observatory, Research Centre for Astronomy and Earth Sciences, Hungarian Academy of Sciences, Konkoly-Thege Mikl{\'o}s {\'u}t 15-17, 1121 Budapest, Hungary}

\author{Thomas~Henning}
\affiliation{Max Planck Institute for Astronomy, K{\"o}nigstuhl 17, 69117 Heidelberg, Germany}

\author{\'{A}gnes~K\'{o}sp\'{a}l}
\affiliation{Konkoly Observatory, Research Centre for Astronomy and Earth Sciences, Hungarian Academy of Sciences, Konkoly-Thege Mikl{\'o}s {\'u}t 15-17, 1121 Budapest, Hungary}
\affiliation{Max Planck Institute for Astronomy, K{\"o}nigstuhl 17, 69117 Heidelberg, Germany}

\author{Eduard~Vorobyov}
\affiliation{Department of Astrophysics, The University of Vienna, Vienna, A-1180, Austria}
\affiliation{Research Institute of Physics, Southern Federal University, Stachki 194, Rostov-on-Don, 344090, Russia}

\author{Dmitri~Wiebe}
\affiliation{Institute of Astronomy, Russian Academy of Sciences, 48 Pyatnitskaya St., Moscow, 119017, Russia}

\begin{abstract}
The FU~Ori-type young stellar objects are characterized by a sudden increase in luminosity by 1--2 orders of magnitude, followed by slow
return to the pre-outburst state on timescales of $\sim$10--100\,yr. The outburst strongly affects the entire 
disk, changing its thermal structure and radiation field. In this paper, using a detailed physical-chemical model we study 
the impact of the FU~Ori outburst on the disk chemical inventory. Our main goal is to identify gas-phase molecular tracers of the outburst activity
that could be observed after the outburst with modern telescopes such as ALMA and NOEMA. We find that the majority of molecules experience a considerable increase 
in the total disk gas-phase abundances due to the outburst, mainly due to the sublimation of their ices. 
Their return to the pre-outburst chemical state takes different amounts of time, from nearly instantaneous to very long. 
Among the former ones we identify CO, NH$_3$, C$_2$H$_6$, C$_3$H$_4$, etc. Their abundance evolution tightly follows the luminosity curve. For CO the abundance increase does not exceed an order of magnitude, while for other tracers the abundances increase by 2--5 orders of magnitude. Other molecules like H$_2$CO and NH$_2$OH have longer retention timescales, remaining in the gas phase for $\sim 10-10^3$\,yr after the end of the outburst. Thus H$_2$CO could be used as an indicator of the previous outbursts in the post-outburst FU~Ori systems. We investigate the corresponding time-dependent
chemistry in detail and present the most favorable transitions and ALMA configurations for future observations. 
\end{abstract}


\section{Introduction}

The so-called luminosity problem in the low-mass star formation theory stems from the apparent discrepancy between the predicted luminosities of young stellar objects (YSOs) and their an order of magnitude lower observed values. One of the possible explanations of this discrepancy is a scenario where the YSO luminosity varies considerably, alternating between short, strong outbursts and prolonged periods of low luminosity states \citep{1996ARA&A..34..207H,2012ApJ...747...52D}. FU~Ori type objects (FUors), which experience $\sim 100 L_{\odot}$ luminosity outbursts, may present an observational manifestation of this phenomenon \citep{2011ARA&A..49...67W,2014prpl.conf..387A}. The class of FUors comprises both Class~I objects with embedding envelopes and the ones with tenuous envelopes close to Class~II YSOs \citep{2007ApJ...668..359Q}. EX~Lup type objects (EXors) that are similar to FUors, but characterized by weaker ($\sim 10 L_{\odot}$) and shorter outbursts, belong usually to later evolutionary stages, although the distinction between these two types is not always clear \citep{2014prpl.conf..387A}.

While astronomers have identified only a handful of EXors and FUors, many more YSOs that are currently in a low luminosity state could have 
experienced similar outbursts in more distant past, pre-dating a modern epoch of accurate photometric measurements. Thus, it is desirable to
find other ways to discern whether an YSO could have experienced the luminosity outbursts, and when the last event could have approximately happened.

Here chemistry comes to rescue, as the luminosity outbursts should strongly affect the disk physical structure and thus leave their footprints on the disk chemical composition \citep{2007JKAS...40...83L,2012ApJ...754L..18V,2012ApJ...758...38K,2013A&A...557A..35V,Rab17}. Intriguingly, \citet{2013ApJ...779L..22J} have found observational signatures of a recent outburst in the Class~0 IRAS~15398--3359 object by the absence of HCO$^+$ emission towards the center of the object,
which was destroyed by the water evaporated from the dust grains during the outburst. It is also conceivable that the changes in the disk 
physical properties due to the outburst may activate alternative chemical pathways, e.g., reactions with barriers or thermal processing of ices.
The impact of the FUori-outburst on the disk and envelope chemical structure and the CO emission has been studied in detail by \citet{Rab17}. A similar role in identifying past outbursts could also be played by mineralogy, as shown by \citet{2009Natur.459..224A} for an EXor type object. The authors reported the in-situ crystallization of amorphous silicate grains in the inner disk of EX Lup during its large outburst in 2008. Follow-up observations, however, revealed that the freshly formed crystals have been gradually disappearing \citep{2012ApJ...744..118J}, suggesting that a measurable drop in the degree of crystallinity in a protoplanetary disk may also signal a recent eruption.

It has been demonstrated that volatiles, such as water, carbon monoxide, carbon dioxide etc., are particularly sensitive to the disk thermal structure 
and its changes \citep{2013A&A...557A..35V,2014FaDi..169...49P,2014prpl.conf..363P,2011ARA&A..49..471M}. While ices evaporate
almost instantaneously due to the outburst-driven temperature increase, their freeze-out back to the dust grains after the outburst fades can be slow.
This makes the respective snow lines to be located farther away than possible in the quiescent disk. 
For instance, \citet{2016Natur.535..258C} observed an abrupt change in the optical depth at 40~au in the disk around FU~Ori type star V883~Ori, which can possibly be explained as a consequence of water snowline shifted outwards by ongoing luminosity outburst. The observed shifts of the CO snow lines 
in the very low luminosity objects (VELLOs) have been reported by \citet{2018arXiv180104524H}, which can also be interpreted as a sign of the past outbursts. 
The outbursts can also affect the dust properties in the disks and hence be indirectly traced by the dust observations. 
Apart from crystallization of silicates \citep{2009Natur.459..224A}, due to an outburst dust grains can be stripped of their ice mantles, which changes their collisional properties and slows down dust growth \citep{Hubbard17}. Besides, if the dust particles grew to the aggregates bound together by the water ice, the outburst can lead to their destruction into smaller monomers, resulting in a sudden decrease of the average dust sizes inside the outburst-driven water snow line \citep{2010A&A...513A..79B,2016Natur.535..258C}.

In this paper we aim to investigate in detail the chemical effects of the luminosity outburst on an FUori-like system at a later Class~II stage of the evolution, which is more easily accessible via (sub-)millimetre observations. This is the follow-up study of our previous work by Wiebe et~al.~(2018; 
submitted to MNRAS), where the impact of the FUori outburst has been studied only in a few selected locations in a disk. 
Here we seek to more completely understand the outburst-induced chemical evolution using the time-dependent 2D spatially resolved calculations. 
The adopted quasi-stationary axisymmetric disk model describes a Class~II system without or with a negligible envelope. 
Various disk masses, sizes and dust size distributions are considered. We use the total disk abundances to isolate potential molecular tracers 
of the outburst and study how their spatial distributions change due to combined physical and chemical effects. Finally,
we use a realistic model based on V346~Nor and perform line radiative transfer calculations to obtain synthetic spectra of the 
most convenient observational outburst tracers.

\section{Model}

\subsection{Disk Structure}
\label{sec:maths} 

Using our ANDES code \citep{Akimkin13}, we perform astrochemical modeling of a protoplanetary disk experiencing a FUor-like luminosity outburst.
The disk structure is similar to the model described in \citet{Molyarova17}. The disk is assumed to be azimuthally symmetric with surface density distribution described by a tapered power-law. The vertical structure is derived assuming hydrostatic equilibrium. The vertical and radial temperature structure 
is computed using a combination of the radiation transfer in the disk atmosphere and a parametric approach in the disk midplane 
\citep{WB2014,isella2016ringed,Molyarova17}. The midplane temperature $T_{\rm m}(r)$ is determined by the combined stellar and accretion luminosities. The temperature in the atmosphere $T_{\rm a}(r,z)$ is calculated using a 2D UV radiation transfer. Density and temperature distributions are self-consistent and calculated iteratively. Dust optical properties are taken from \citet{1993ApJ...402..441L}. A total dust mass is assumed to be $1\%$ of the gas mass. 
The adopted  dust size distribution is described in Section~\ref{sec:chemistry}.

A total luminosity of the source consists of an intrinsic stellar luminosity and an accretion luminosity that varies as the outburst proceeds 
(see Section~\ref{sec:outburst} for details). The vertical hydrostatic disk structure is recalculated as the total luminosity varies. Here we neglect possible hydrodynamical effects of the outburst on the disk structure and assume that it attains the equilibrium instantaneously with the luminosity change.

According to recent millimeter observations, the FUori disks may be quite massive and compact \citep{2018A&A...612A..54L,2018MNRAS.474.4347C}, with masses up to $10-20\%$ of the stellar masses and radii of tens of au. Thus, we consider a range disk masses $M_{\rm disk} = 0.01, 0.1\,M_{\odot}$ and characteristic radii $R_{\rm c}=100$~and $20$\,au. In addition, two dust size distributions referred to as ``medium'' and ``grown'' dust models are considered 
(see Section~\ref{sec:chemistry} for more details). In total, our main computational set includes eight disk models. 
Additional models with a smaller disk mass $M_{\rm disk} = 0.001 M_{\odot}$ are also considered but only used for the chemical analysis of some 
peculiar species. Our reference disk model has a characteristic radius $R_{\rm c} = 100$\,au, a mass $M_{\rm disk} = 0.01 M_{\odot}$, and the 
``medium'' dust size distribution.

\begin{figure*}
\includegraphics[width=2.0\columnwidth]{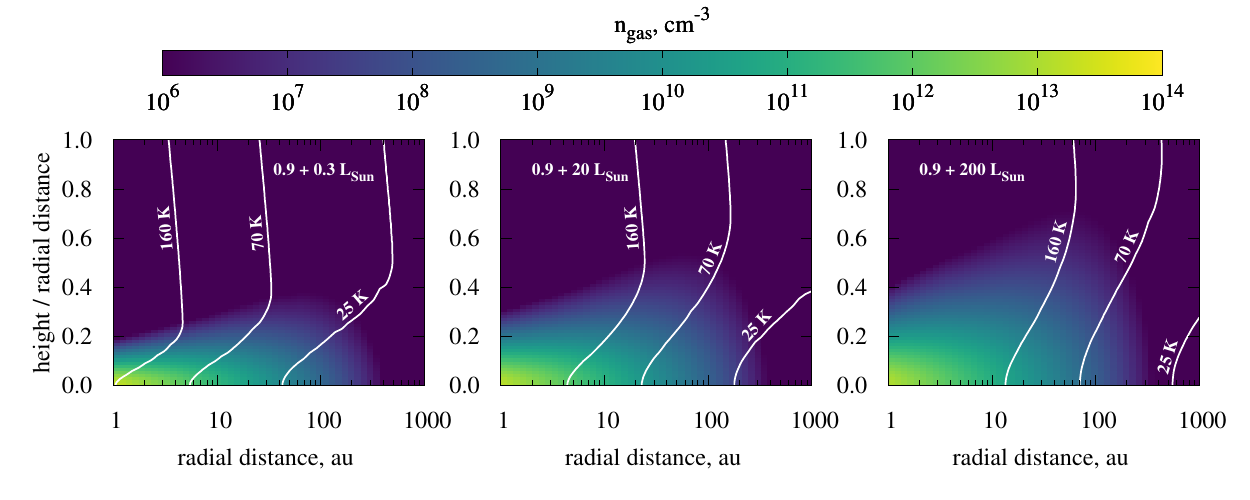}
\caption{The gas number density (color) and temperature distribution (isolines) in the reference disk model ($R_{\rm c} = 100$\,au, $M_{\rm disk} = 0.01 M_{\odot}$, ``medium'' dust size) at different epochs of the outburst. The stellar luminosity is a constant equal to $0.9 L_{\odot}$. The accretion
luminosity $L_{\rm acc}$ is variable. The left panel shows the disk structure when $L_{\rm acc}=0.3 L_{\odot}$ (quiescent stage), the middle panel is for the case when $L_{\rm acc}=20 L_{\odot}$ (intermediate state during luminosity rise or decrease), and the right panel is for the case of the maximum luminosity with $L_{\rm acc}=200 L_{\odot}$ (outburst maximum). The temperature isolines mark approximate condensation fronts of the main disk volatiles: 25\,K for CO, 70\,K for CO$_2$, and 160\,K for H$_2$O, respectively \citep{2015A&A...582A..41H}.}
  \label{fig:diskstruct}
\end{figure*}

Figure~\ref{fig:diskstruct} shows the computed disk physical structure for the reference model at different time moments: 1) at a 
quiescent state, when $L_{\rm tot}=L_{\rm star}+L_{\rm acc}=0.9 + 0.3\,L_{\odot}$ (Figure~\ref{fig:burstprof}, time moments 1 and 3), 2) during the luminosity rise, when $L_{\rm tot}=0.9 + 20\,L_{\odot}$, and 3) at the maximum of the luminosity, when $L_{\rm tot}=0.9 + 200\,L_{\odot}$ (Figure~\ref{fig:burstprof}, time moment 2). As can be clearly seen, as the accretion luminosity increases, the disk 
heats up and gets thicker and more flaring. At the maximum luminosity of $200\,L_{\odot}$, the disk midplane temperature rises above 50\,K even at 
100\,au distance from the star. Note, however, that our models are quasi-stationary, while during a typical outburst ($\sim100$ yr) not all disk regions may have enough time to adapt to the increased irradiation and reach an equilibrium state (for the relevant thermal and dynamical timescales in a disk, see e.g. Eqs.~24--30 in \citet{1997ApJ...490..368C} or Fig.~3 in~\citet{2014ARep...58..522V}).

\subsection{Chemistry and grain properties}
\label{sec:chemistry}
We also employ ANDES to calculate the evolution of disk chemical composition. The ANDES chemical engine is based on the modified ALCHEMIC
network \citep{2011ApJS..196...25S}, which includes an extended updated set of gas-phase and surface reactions from \citet{2015ApJS..217...20W}. The adopted chemical network comprises 650 species involved in 7807 reactions.
Various surface processes, such as the 1\% chemical reactive desorption, hydrogen tunneling through reaction barriers, high-energy photodesorption and 
photoprocessing of ices (with UV desorption yield for ices treated as in \citet{1999A&A...342..542W}) are considered. 
The cosmic rays, X-rays, and the decay of short-lived radioactive elements are considered as the main sources of ionization. 
The chemical structure changes during the evolution from a protostellar core to a disk phase, as the material passes through varying physical conditions \citep{2009A&A...495..881V,2016MNRAS.462..977D,2016A&A...595A..83E}. Being unable to reproduce such effects in a quasi-stationary disk model, we restrict our 
approach by adopting the composition of a 1\,Myr old molecular cloud as the disk initial chemical composition. 
It is calculated using a single-point model with a density of $10^{4}$\,cm${^{-3}}$, a temperature of 10\,K, no UV radiation field, and 
an average grain size of 0.1\,$\mu$m dust, starting from the ``low-metals'' elemental abundances of \citet{1998A&A...334.1047L}. 
This initial disk chemical composition is quite diverse and includes complex species in both the gas and ice phases. 

The adopted dust size distribution $f(a)$ is based on the MRN model \citep{1977ApJ...217..425M} with various maximum grain radii $a_{\rm max}$. 
These dust distributions are used to calculate the radiation field and temperature in the disk atmosphere. For surface chemistry the only dust 
parameter that matters is the dust surface area (per unit gas volume), which is calculated using the adopted dust size distributions.

We consider the following two cases of the dust population: the so-called ``medium'' dust with $a_{\rm max}=2.5\times10^{-3}$\,cm and 
the ``grown'' dust with $a_{\rm max}=2.5\times10^{-1}$\,cm. The minimum grain size $a_{\rm min}=5\times10^{-7}$\,cm is kept the same in both models. 
The dust size distribution is identical in each point of the disk (no settling, no radial drift). 
The average grain radii for these two dust populations can be calculated as \citep{2011ApJ...727...76V}
\begin{equation}
\overline{a} = \left. \int\limits_{a_{\rm min}}^{a_{\rm max}} f(a) a^3 da \middle/ \int\limits_{a_{\rm min}}^{a_{\rm max}} f(a) a^2 da \right.,
\label{eq:mean}
\end{equation}
which gives $\overline{a}=3.7\times10^{-5}$ and $\overline{a}=3.7\times10^{-4}$\,cm, respectively. Inside the chemical model, reaction rates are calculated using total surface area of dust particles ensemble, consisting of grains with the size $\overline{a}$.

\subsection{Outburst}
\label{sec:outburst}

\begin{figure}
\includegraphics[width=\columnwidth]{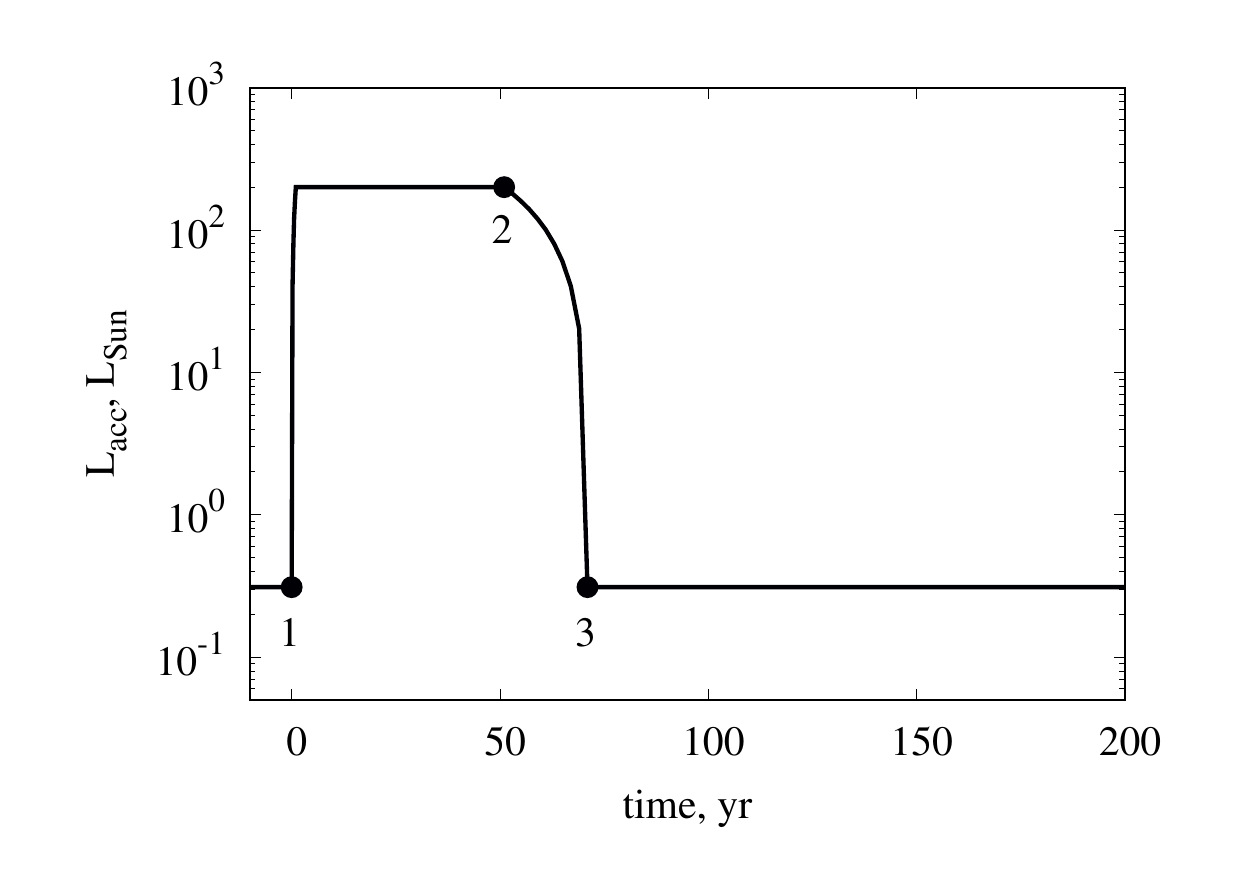}
 \caption{The accretion luminosity during the outburst. The initial moment $t=0$ is the last time moment of the quiescent phase before the outburst.
 Points 1--3 mark $t=0$ (time moment 1), $51$ (time moment 2), and $71$ (time moment 3) years after the beginning of the outburst, respectively.}
  \label{fig:burstprof}
\end{figure}

The initial chemical inventory at the outset of the disk stage corresponds to the 1\,Myr evolution of a molecular cloud core as described above. 
Simultaneous modeling of the disk formation and non-equilibrium chemistry is a computationally challenging task \citep{2013A&A...557A..35V}. 
In this work we restrict ourselves to a simple treatment of the pre-outburst disk evolution by running the chemistry for 0.5\,Myr, assuming 
quiescent disk physical conditions \citep[see][for the discussion on the length of pre-outburst phase]{W18}. 
During all the stages the stellar luminosity is constant at $0.9\,L_{\odot}$, which roughly corresponds to a young solar mass star. 
The quiescent accretion luminosity is $0.3\,L_{\odot}$, assuming the quiescent accretion rate of $10^{-8}\,M_{\odot}$\,yr$^{-1}$. 
Figure~\ref{fig:burstprof} shows the assumed outburst luminosity profile. The outburst is simulated by increasing the accretion luminosity 
up to $200\,L_{\odot}$ (corresponding to the peak accretion rate of $\sim10^{-5}\,M_{\odot}$\,yr$^{-1}$). 
The luminosity rises linearly within 1\,yr, then remains constantly high for 50\,yr, and after that drops linearly to the 
initial quiescent state value for 20\,yr. We follow the disk chemical evolution for 500\,000\,yr before the outburst,
then for 71~yr during the outburst, and then for another 10\,000\,yr after the outburst.

The X-ray temporal behavior during the FUors outbursts is currently not well understood, with some detections and non-detections 
\citep{2004Natur.430..429K,2009ApJ...696..766S,2014A&A...570L..11L}. At the same time, non-FUor objects can show outburst activity in X-ray, e.g., V\,1647~Ori \citep{2010ApJ...714L..16H,2011ApJ...741...83T,2012ApJ...754...32H}. In this work we assume the constant X-ray luminosity of 
$10^{30}$\,erg\,s$^{-1}$ during the outburst and quiescent phases. This assumption is supported by the observations of variable YSOs  \citep{2010ApJ...722.1654S,2010ATel.3040....1P,2015ATel.7025....1P,2016ATel.8548....1A}.
UV intensities are calculated as black-body radiation from the accretion region, which has a constant temperature of $20000$\,K and a size scaling with the luminosity growth.

The luminosity outburst causes rise of the disk temperature and radiation intensity and also changes the vertical density structure, as in our model 
it is defined self-consistently by iterations between thermal, density, and radiation field distributions. 
While recalculating disk physical structure, we keep relative abundances constant at every disk cell.

\section{Results}

In this section we analyze the effect of the luminosity outburst on the abundances of various chemical species in the disk. 
We identify species sensitive to the luminosity rise and also search for long-term outburst tracers, that is, the molecules that retain 
elevated gas-phase abundances long after the disk has returned to the quiescent state. 
We examine the influence of the disk physical properties and dust distribution on the behavior of the most promising tracers.

\subsection{CO abundance}
\label{sec:COab}

\begin{figure}
\includegraphics[width=\columnwidth]{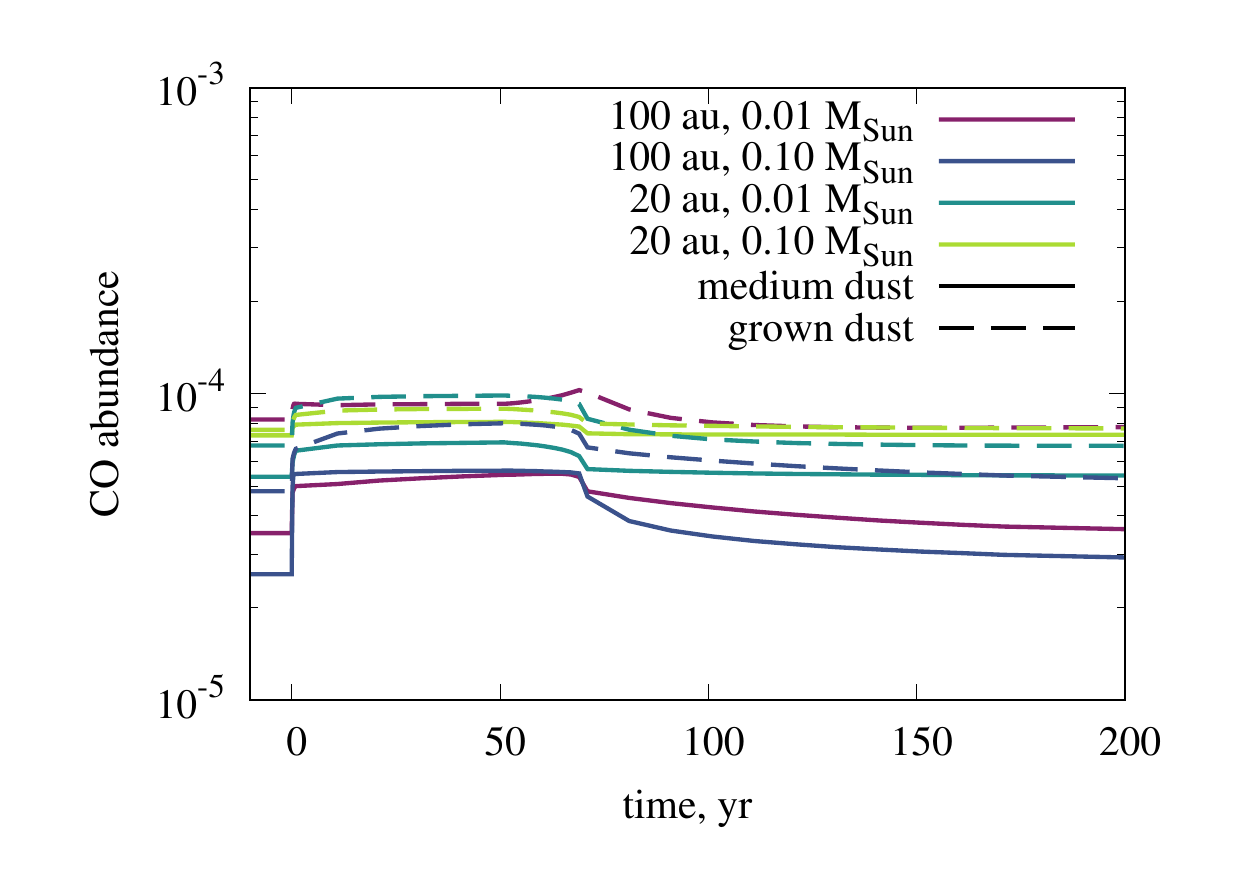}
 \caption{The average gas-phase CO abundances in different disk models. 
 Solid lines are for the models with ``medium'' dust, dashed lines are for the models with ``grown'' dust.}
  \label{fig:COmass}
\end{figure}

First, we consider CO as one of the most abundant and readily observed molecules, which has been suggested to bear an imprint of the outburst 
long after its end in the Class~I systems \citep{Rab17}. We find that the total amount of CO in the disk is sensitive to the disk properties. 
It is relatively well correlated with the disk mass, however the CO abundance is significantly lower than the interstellar value of $\sim 10^{-4}$ (wrt H) due to both freeze-out and surface chemistry~\citep{Molyarova17}. As the outburst commences, CO is evaporated from the dust surfaces. The conversion of the CO ice into the less volatile CO$_2$ ice during the quiescent stage may suppress the release of CO to the gas phase during the outburst. 

Figure~\ref{fig:COmass} shows the change in the total, disk-integrated abundance of gaseous CO due to the outburst in the disk models that differ in mass, 
characteristic radius, and dust size distribution. The total CO abundances differ between the models by about a factor of 3 
even before the onset of the outburst. The magnitude of these differences is of the same order as the change due to the outburst itself. 
Thus, the CO abundance in the disk demonstrates variability that is not only due to the episodic heating event 
but are also determined by the disk physics. We can conclude that in the case of an evolved FUori disk without an envelope CO is not a reliable tracer 
of the past outburst activity. 

CO molecule could, however, serve as a reference molecule in this case, as an order of magnitude variability is not actually very high. If we have some other species as an outburst tracer and measure its high molecular signal, it may indicate either a high disk mass or an unusually high abundance caused by past outburst. A comparison with CO, using the abundance ratios of the given molecule to CO, would break this degeneracy. 

Other studies \citep{2013A&A...557A..35V,Rab17} showed that CO is able to stay in the gas phase for centuries after the outburst, much longer than we see 
in our simulations. This difference is explained by the absence of a low-density envelope with longer chemical and freeze-out timescales in 
our model. 

To provide a more detailed comparison, we run a test model of a disk with the same parameters as in \citet{Rab17}. The purpose is to compare the effect of different chemical models. Apart from chemical network differences, there are also differences in the outburst modeling. Thus, not expecting to get exactly the same result, we intend to show the qualitative differences between the models.

The main difference is that considering more evolved objects, we do not take into account the embedding envelope. However, the disk model is very similar, being slightly different in density and temperature distributions, but with identical surface density profile. Unlike in \citet{Rab17}, the vertical density distribution in our model depends on the current luminosity, thus evolving during the outburst. The dust size distribution is the same everywhere in the disk with minimum and maximum values of 0.005~and 1000\,$\mu$m, we adopt the same parameters for the benchmarking. While \citet{Rab17} additionally have a smaller dust population with $a_{\rm max}=1$\,$\mu$m  characteristic of the envelope, which we do not include in our model. We consider a smaller disk region of only 1000~au from the star. The outburst profile is somewhat different as well, as described in Section~\ref{sec:outburst}, having smoother edges instead of being sharp with sudden luminosity increase and drop. The duration of the phases in the present model are $5\times10^5$ instead of $10^5$\,yr for the quiescent stage and 70 instead of 100\,yr for the outburst stage. The outburst magnitude is two times larger in our model.

\begin{figure}
\includegraphics[width=\columnwidth]{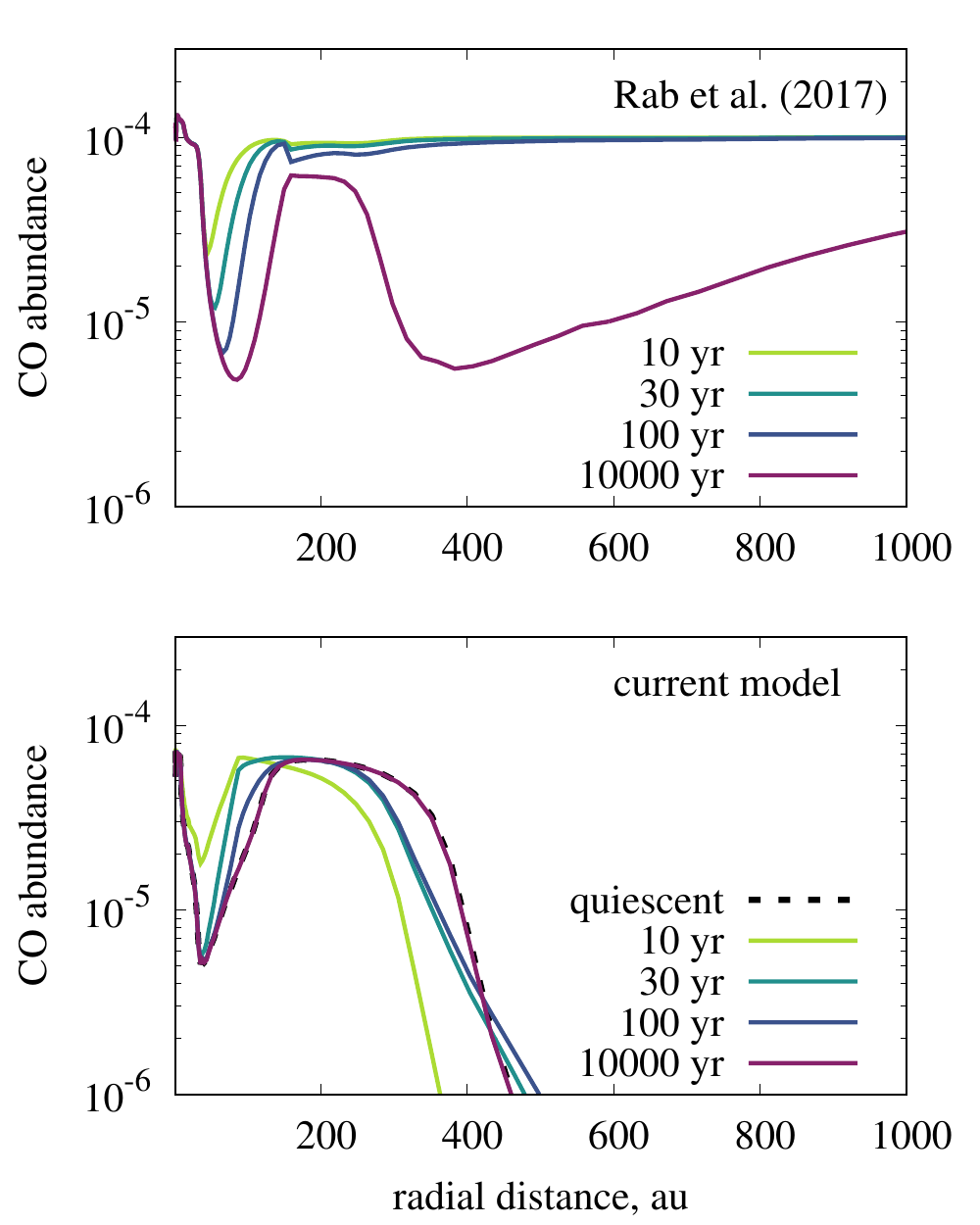}
 \caption{Vertically integrated abundance of gaseous CO before and some time (as denoted by graph legend) after the outburst. Upper panel is ProDiMo results, bottom panel is ANDES results. Differences at outer radii are due to the envelope present in ProDiMo model.}
  \label{fig:rabcomparison}
\end{figure}

Figure~\ref{fig:rabcomparison} shows the radial distribution of gas-phase CO in the test model along with the analogous distribution calculated with \texttt{ProDiMo} from \citet{Rab17}. 

We find that in general our results are similar, albeit with recognizable discrepancies. The models share CO depletion area between 50~and 150~au, although it is shifted inwards in our model due to somewhat colder thermal structure. The shift is possibly caused by CO-ice conversion into CO$_2$-ice, which effectively depletes CO in the adopted model \citep[see][]{Molyarova17}. After the outburst the profile returns to the quiescent state with approximately the same rate in both models. Timescales of CO abundance change after the outburst are faster in our model, due to the absent population of small cold dust associated with the envelope, which sustains slow surface chemistry.

In the outer disk ($R > 200$\,au), \texttt{ProDiMo} CO abundance decreases, while in our model it is growing. In the first case the decrease is explained by CO freeze-out in the envelope. In our model the envelope is absent, so the UV radiation field is high, and CO is effectively photodissociated. With the luminosity growth the destruction rate increases, diminishing CO abundance, and after the outburst it is gradually restored under milder UV.

The main difference comes from the lacking envelope, which results in a very low CO abundance beyond 400~au, where CO is photo-dissociated under unscreened interstellar radiation field. The profile corresponding to $10^4$\,yr after the outburst is nearly the same as the quiescent profile before the outburst.

\subsection{Molecular tracers of the outburst}
Unlike carbon monoxide that shows only a factor of a few variations in its abundance due to the outburst, 
other molecules can be more sensitive. We split the possible outburst tracers into two groups. The first group includes species whose abundances
are only sensitive to the ongoing outburst and quickly return to the quiescent phase after the outburst  ceases. The second group is comprised of 
species which take longer time to return to the pre-outburst phase or which remain overabundant after the outburst. 
We discuss the first group in Section~\ref{sec:immediate} and the second group in Section~\ref{sec:postburst}.

\subsubsection{Immediate tracers}
\label{sec:immediate}

As a first step to sort out promising outburst tracers, we analyze the total abundances of all species and their evolution during the outburst. The further analysis of the 2D spatial distributions of selected species is presented in Section~\ref{sec:2d}.

Naturally, there are species that are very sensitive to the outburst, with their abundances changing immediately with the rise of the luminosity. 
Majority of the gas-phase species experience an increase in concentration, predominantly due to their evaporation from dust grains. 
To classify the molecules into different categories, we compare their total disk abundances during the outburst $A_{2}$ at the time moment `2' (Figure~\ref{fig:burstprof}) with their pre-outburst values $A_{1}$ (at the time moment `1'). As immediate tracers, we only consider species whose 
outburst abundances $A_{2}$ exceed $10^{-9}$, with a strong abundance increase of $\log_{10}(A_{2}/A_{1}) > 2$
in more than half of the disk models, see Table~\ref{tab:burstmarkers} (upper part). The results for the two dust populations are presented separately. 

\begin{table*}
  \centering  
  \caption{Immediate tracers of the outburst (upper part) and species observed in disks (bottom part), sorted by their abundance in the reference model. Ranges shown encompass data for all considered disk masses and sizes. Species whose outburst abundances are determined by the gas-phase chemistry are marked with $\bigcirc$, while those whose abundances change mainly due to the evaporation of ices are marked with $\ast$ ($\ast$ $\bigcirc$ means that both processes are important, either in different disk regions or in different models). $A_{1}$ is the total pre-outburst species abundance in disk, $A_{2}$ is the abundance during the outburst. Shown ranges are for disk models with various disk masses and radii.}
  \label{tab:burstmarkers}
  \begin{tabular}{rllrllr}
   & Species name & $A_{2}$ & $\log_{10}(A_{2}/A_{1})$ & & $A_{2}$ & $\log_{10}(A_{2}/A_{1})$ \\ 
    &            & medium dust &                & & grown dust &                 \\   
    \hline
 $\ast$            & NH$_3$        & $  2.6 \times 10^{-06} \dots 1.2 \times 10^{-05}$  & $ 2.2 \dots 2.5$ & & $  1.1 \times 10^{-06} \dots 1.3 \times 10^{-05}$ & $ 1.7 \dots 2.0$ \\ 
 $\ast$            & H$_2$CO       & $  1.3 \times 10^{-06} \dots 1.0 \times 10^{-05}$  & $ 4.5 \dots 5.7$ & & $  1.3 \times 10^{-07} \dots 1.4 \times 10^{-05}$ & $-0.5 \dots 5.8$ \\
 $\ast$            & C$_3$H$_4$    & $  1.5 \times 10^{-06} \dots 3.7 \times 10^{-06}$  & $ 2.9 \dots 4.4$ & & $  2.5 \times 10^{-07} \dots 1.7 \times 10^{-06}$ & $ 2.3 \dots 3.2$ \\
 $\ast$            & C$_2$H$_6$    & $  6.8 \times 10^{-07} \dots 2.3 \times 10^{-06}$  & $ 2.2 \dots 3.3$ & & $  2.7 \times 10^{-07} \dots 1.6 \times 10^{-06}$ & $ 1.9 \dots 3.3$ \\                                                       
 $\ast$ $\bigcirc$ & CH$_3$OH      & $  1.0 \times 10^{-07} \dots 1.4 \times 10^{-06}$  & $ 3.7 \dots 3.9$ & & $  4.5 \times 10^{-08} \dots 1.5 \times 10^{-06}$ & $ 3.1 \dots 4.3$ \\
 $\ast$ $\bigcirc$ & HCOOH         & $  3.0 \times 10^{-08} \dots 1.4 \times 10^{-07}$  & $ 3.3 \dots 3.6$ & & $  2.7 \times 10^{-08} \dots 1.5 \times 10^{-07}$ & $ 3.2 \dots 3.8$ \\
 $\ast$            & NH$_2$OH      & $  1.8 \times 10^{-09} \dots 3.4 \times 10^{-08}$  & $ 4.9 \dots 5.4$ & & $  5.4 \times 10^{-09} \dots 3.4 \times 10^{-08}$ & $ 4.9 \dots 6.2$ \\
 $\ast$            & CH$_3$CHO     & $  2.3 \times 10^{-08} \dots 7.2 \times 10^{-08}$  & $ 5.1 \dots 5.7$ & & $  2.5 \times 10^{-10} \dots 1.1 \times 10^{-07}$ & $ 3.7 \dots 6.1$ \\                                                       
 $\ast$            & H$_5$C$_3$N   & $  4.7 \times 10^{-09} \dots 2.1 \times 10^{-08}$  & $ 2.8 \dots 3.3$ & & $  2.0 \times 10^{-09} \dots 1.2 \times 10^{-08}$ & $ 2.7 \dots 2.9$ \\
 $\ast$ $\bigcirc$ & C$_6$H$_4$    & $  2.6 \times 10^{-09} \dots 1.5 \times 10^{-08}$  & $ 3.0 \dots 3.7$ & & $  2.5 \times 10^{-09} \dots 1.6 \times 10^{-08}$ & $ 2.7 \dots 3.2$ \\ 
        $\bigcirc$ & HC$_5$N       & $  4.1 \times 10^{-09} \dots 1.8 \times 10^{-08}$  & $ 4.4 \dots 5.0$ & & $  1.9 \times 10^{-12} \dots 8.4 \times 10^{-09}$ & $ 2.6 \dots 4.3$ \\                                                      
 $\ast$            & CH$_3$OCH$_3$ & $  3.8 \times 10^{-09} \dots 2.4 \times 10^{-08}$  & $ 3.1 \dots 6.3$ & & $  1.7 \times 10^{-11} \dots 3.4 \times 10^{-08}$ & $ 2.5 \dots 3.3$ \\                                                      
        $\bigcirc$ & HC$_7$N       & $  3.7 \times 10^{-09} \dots 2.0 \times 10^{-08}$  & $ 5.6 \dots 6.4$ & & $  3.1 \times 10^{-13} \dots 9.2 \times 10^{-09}$ & $ 3.3 \dots 6.1$ \\                                                      
 $\ast$            & H$_2$S$_2$    & $  2.1 \times 10^{-09} \dots 8.0 \times 10^{-09}$  & $ 2.6 \dots 3.3$ & & $  1.0 \times 10^{-09} \dots 5.5 \times 10^{-09}$ & $ 1.7 \dots 3.3$ \\                                                      
 $\ast$ $\bigcirc$ & CH$_5$N       & $  1.9 \times 10^{-09} \dots 8.0 \times 10^{-08}$  & $ 2.3 \dots 3.4$ & & $  1.3 \times 10^{-09} \dots 4.7 \times 10^{-08}$ & $ 2.1 \dots 2.8$ \\
        $\bigcirc$ & HC$_3$N       & $  1.8 \times 10^{-09} \dots 8.0 \times 10^{-09}$  & $ 2.3 \dots 2.8$ & & $  1.2 \times 10^{-10} \dots 5.9 \times 10^{-09}$ & $ 0.7 \dots 2.0$ \\
 $\ast$            & HCOOCH$_3$    & $  7.8 \times 10^{-10} \dots 1.6 \times 10^{-08}$  & $ 6.6 \dots 8.4$ & & $  4.4 \times 10^{-10} \dots 1.6 \times 10^{-08}$ & $ 6.7 \dots 7.6$ \\
 \hline                                             
 $\ast$ $\bigcirc$ & CO            & $  5.4 \times 10^{-05} \dots 8.1 \times 10^{-05}$  & $ 0.0 \dots 0.3$ & & $  8.0 \times 10^{-05} \dots 9.9 \times 10^{-05}$ & $ 0.0 \dots 0.2$ \\
 $\ast$            & CO$_2$        & $  3.3 \times 10^{-05} \dots 4.6 \times 10^{-05}$  & $ 0.9 \dots 1.4$ & & $  8.4 \times 10^{-06} \dots 3.8 \times 10^{-05}$ & $-0.1 \dots 0.9$ \\
 $\ast$            & H$_2$O        & $  2.0 \times 10^{-05} \dots 1.0 \times 10^{-04}$  & $ 0.8 \dots 2.4$ & & $  1.4 \times 10^{-05} \dots 1.1 \times 10^{-04}$ & $ 1.0 \dots 2.0$ \\
        $\bigcirc$ & O             & $  1.4 \times 10^{-05} \dots 1.4 \times 10^{-04}$  & $ 0.4 \dots 0.9$ & & $  3.0 \times 10^{-05} \dots 2.3 \times 10^{-04}$ & $ 0.2 \dots 0.7$ \\
        $\bigcirc$ & C$^+$         & $  3.1 \times 10^{-07} \dots 2.9 \times 10^{-05}$  & $ 0.9 \dots 1.1$ & & $  7.3 \times 10^{-07} \dots 5.1 \times 10^{-05}$ & $ 0.7 \dots 0.8$ \\
 $\ast$            & HCN           & $  1.4 \times 10^{-07} \dots 2.4 \times 10^{-07}$  & $ 1.0 \dots 1.6$ & & $  3.1 \times 10^{-08} \dots 1.7 \times 10^{-07}$ & $ 0.5 \dots 1.7$ \\
        $\bigcirc$ & C$_2$H$_2$    & $  1.3 \times 10^{-08} \dots 4.8 \times 10^{-08}$  & $ 1.8 \dots 2.3$ & & $  8.7 \times 10^{-09} \dots 4.7 \times 10^{-08}$ & $ 1.3 \dots 1.7$ \\
 $\ast$            & HNC           & $  9.4 \times 10^{-09} \dots 2.0 \times 10^{-08}$  & $ 1.3 \dots 1.9$ & & $  1.5 \times 10^{-09} \dots 8.8 \times 10^{-09}$ & $ 0.8 \dots 1.8$ \\
        $\bigcirc$ & SO            & $  8.0 \times 10^{-09} \dots 1.7 \times 10^{-08}$  & $-0.1 \dots 0.3$ & & $  4.6 \times 10^{-09} \dots 1.6 \times 10^{-08}$ & $-0.5 \dots-0.1$ \\
        $\bigcirc$ & OH            & $  6.4 \times 10^{-09} \dots 7.0 \times 10^{-08}$  & $-0.8 \dots 1.7$ & & $  1.2 \times 10^{-08} \dots 7.3 \times 10^{-08}$ & $ 0.1 \dots 2.1$ \\
 $\ast$ $\bigcirc$ & C$_3$H$_2$    & $  3.4 \times 10^{-10} \dots 7.8 \times 10^{-10}$  & $ 1.6 \dots 2.1$ & & $  3.2 \times 10^{-10} \dots 4.4 \times 10^{-09}$ & $ 2.6 \dots 3.7$ \\
        $\bigcirc$ & CS            & $  9.9 \times 10^{-11} \dots 2.2 \times 10^{-10}$  & $-0.8 \dots 0.2$ & & $  1.5 \times 10^{-11} \dots 1.0 \times 10^{-10}$ & $-1.4 \dots 0.1$ \\
        $\bigcirc$ & CN            & $  4.0 \times 10^{-11} \dots 7.0 \times 10^{-09}$  & $ 0.7 \dots 1.0$ & & $  4.0 \times 10^{-11} \dots 1.5 \times 10^{-09}$ & $-0.1 \dots 1.1$ \\
        $\bigcirc$ & C$_2$H        & $  2.5 \times 10^{-11} \dots 1.2 \times 10^{-09}$  & $ 0.4 \dots 1.0$ & & $  2.6 \times 10^{-11} \dots 3.0 \times 10^{-10}$ & $-0.3 \dots 0.7$ \\
        $\bigcirc$ & HCO$^+$       & $  2.9 \times 10^{-12} \dots 2.6 \times 10^{-11}$  & $-1.2 \dots-0.5$ & & $  1.6 \times 10^{-12} \dots 1.5 \times 10^{-11}$ & $-1.3 \dots-0.1$ \\
        $\bigcirc$ & CH$^+$        & $  8.3 \times 10^{-14} \dots 3.7 \times 10^{-12}$  & $ 0.2 \dots 1.3$ & & $  3.7 \times 10^{-13} \dots 1.6 \times 10^{-11}$ & $ 0.4 \dots 1.1$ \\
        $\bigcirc$ & N$_2$H$^+$    & $  8.5 \times 10^{-15} \dots 1.5 \times 10^{-13}$  & $-2.3 \dots-1.3$ & & $  8.7 \times 10^{-16} \dots 8.0 \times 10^{-15}$ & $-2.4 \dots-3.6$ \\ 
                                                                  
  \end{tabular}
\end{table*}

As can be clearly seen from Table~\ref{tab:burstmarkers}, most of these molecules were frozen as ices prior to the outburst and evaporated from the dust surfaces due to the temperature increase (these species are marked with $\ast$), with the exception of the simplest species and cyanopolyynes. The immediate outburst tracers are relatively simple and abundant volatiles, e.g., NH$_3$, H$_2$CO, hydrocarbons, as well as more complex and less abundant organic compounds, like HCOOH, CH$_3$OH, CH$_3$OCH$_3$, cyanopolyynes, etc. Many of the COMs are not yet detected in disks by the IR/radio observations. The monitoring of the disk chemical composition during the FUor/EXor outburst may provide an opportunity to examine the composition of organic ices in disks.

Observable species that are sensitive to the outburst and whose abundances increase are NH$_3$, H$_2$CO, HCOOH, CH$_3$OH, and HC$_3$N. 
Contrary, molecular ions like HCO$^+$ and N$_2$H$^+$, which are destroyed by the gaseous water and CO evaporated from grains, show decline in their concentrations. The sensitivity of HCO$^+$ to the outburst was used by \citet{2013ApJ...779L..22J} to observationally confirm past outburst 
activity in the low-mass protostellar envelopes. Other observed species in disks that do not meet the above criteria, i.e. either underabundant or insensitive to the outburst, are presented in the bottom of Table~\ref{tab:burstmarkers}. Among these, the abundances of C$_2$H$_2$ and C$_3$H$_2$ can strongly increase by up to two orders of magnitude, but only in a few disk models.  

Interestingly, there is no marked difference between various dust models, while the outcome in the models with the same dust population
but different disk masses and radii may vary. The models with grown dust grains have less dust surface per unit gas volume and hence 
have less efficient surface chemistry, with correspondingly smaller amount of ices and less dramatic ice evaporation due to the outburst.

\subsubsection{Post-outburst tracers}
\label{sec:postburst}

To choose the molecules that can trace the outburst after it has finished, we need to assess the timescales at which their abundances return to the quiescent 
phase. There are many possible reasons for species to change their abundances; the molecule may appear in the gas phase at the time of the outburst due to 
the ice sublimation and then it can be frozen out again or be chemically destroyed; the molecule can be photo-dissociated but then re-formed 
after the outburst, etc. 

We describe the post-outburst chemical evolution in terms of the chemical retention timescale $\tau_{\rm ch}$, which we define as a time needed to compensate 
for the half of the outburst-induced abundance change on a logarithmic scale: $\log_{10} A(\tau_{\rm ch})=\log_{10} A_1+(1/2)\log_{10} (A_2/A_1)$. As we measure this time since the start of the luminosity decline, the species which chemical evolution ideally mirrors the luminosity curve should have $\tau_{\rm ch}\approx20$\,yr, which is the time interval we set for the outburst to fade out.

These retention timescales were calculated for each gas-phase species with $\log_{10} (A_2/A_1)$ greater than one in the 8 disk models (two disk masses 0.1~and 0.01\,$M_{\odot}$, two characteristic radii 20~and 100\,au, two dust models). 
As we are mainly interested in potentially detectable species after the outburst, we do not consider rare species and include only species with either 
$A_{1}$ or $A_{2}$ higher than $10^{-9}$ and with retention timescales $\tau_{\rm ch} > 20$~yr. 

\begin{figure}
\includegraphics[width=\columnwidth]{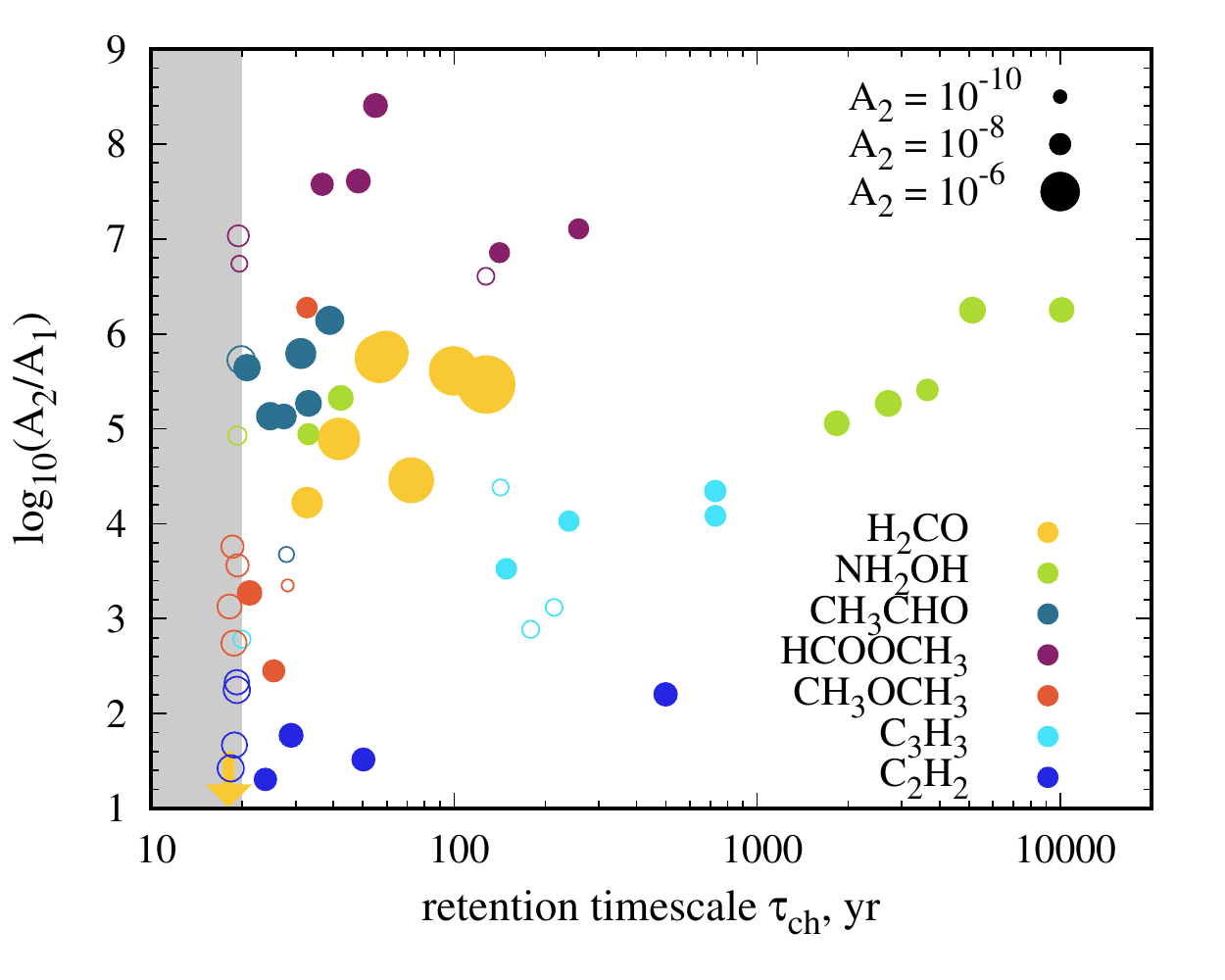}
 \caption{The impact of the outburst on the abundances of selected species $\log_{10}(A_{2}/A_{1})$ and their gas-phase retention timescales $\tau_{\rm ch}$. 
 The circle size corresponds to the abundance value during the outburst ($A_{2}$). Open circles indicate species with either 
 $A_{1}$ or $A_{2}<10^{-9}$ or $\tau_{\rm ch} < 20$~yr. The yellow arrow marks H$_2$CO in the disk model where its abundance decreases during the 
 outburst. The shaded region shows the area of $\tau_{\rm ch} < 20$\,yr, where the majority of species belong.}
  \label{fig:timescales}
\end{figure}

Based on these criteria, we select potential tracers of the past outbursts with high retention timescales $\tau_{\rm ch}$ in majority (more than 4) 
of the considered disk models. These species are H$_2$CO, NH$_2$OH, CH$_3$CHO, CH$_3$OCH$_3$, HCOOCH$_3$, C$_3$H$_3$, and C$_2$H$_2$. In 
Figure~\ref{fig:timescales} we compare their abundance increase during the outburst to their retention timescales. 
Each point corresponds to a specific species in a specific disk model. 

As can be clearly seen, the highest value of $\tau_{\rm ch} \gtrsim 10^{3}$~years is obtained for hydroxylamine (NH$_2$OH). 
The hydroxylamine abundance can grow by 5--6 orders of magnitude, up to the values of $10^{-9}-10^{-8}$, which can make it
potentially detectable in outbursting systems. Similarly, the methyl formate (HCOOCH$_3$) abundance rises by many orders of 
magnitude but remains quite low during the outburst, $\sim 10^{-9}-10^{-8}$. 
Situation is similar for C$_3$H$_3$, which abundance is only relatively high ($>10^{-9}$) during the outburst and is
below $10^{-11}$ soon after the outburst ceases.

Three species with relatively short memory timescales are CH$_3$CHO, CH$_3$OCH$_3$, and C$_2$H$_2$, with $\tau_{\rm ch} \lesssim 20-50$~years
in the majority of disk models. Despite their not very high abundance increase, they can still be present in disks during or right after the outburst. 

The most promising species from observational perspective is formaldehyde (H$_2$CO). Its abundance grows by 4--6 orders of magnitude during the outburst, 
and it takes 30-120~years to deplete formaldehyde back onto the dust grains. The formaldehyde abundance during the outburst in the majority of models 
is also quite high ($\sim10^{-6}$). The only exception is the model marked with an arrow in Figure~\ref{fig:timescales}, where 
the H$_2$CO abundance returns rapidly to the quiescent level and is low during the outburst. This model has a large characteristic radius $R_{\rm c} = 100$\,au and a low mass $M_{\rm disk} = 0.01 M_{\odot}$ (grown dust), and shows no significant presence of the H$_2$CO ice before the outburst. 
Apparently, the surface chemistry in this model is not efficient enough to produce sufficient amount of H$_2$CO before the outburst.

\begin{figure}
\includegraphics{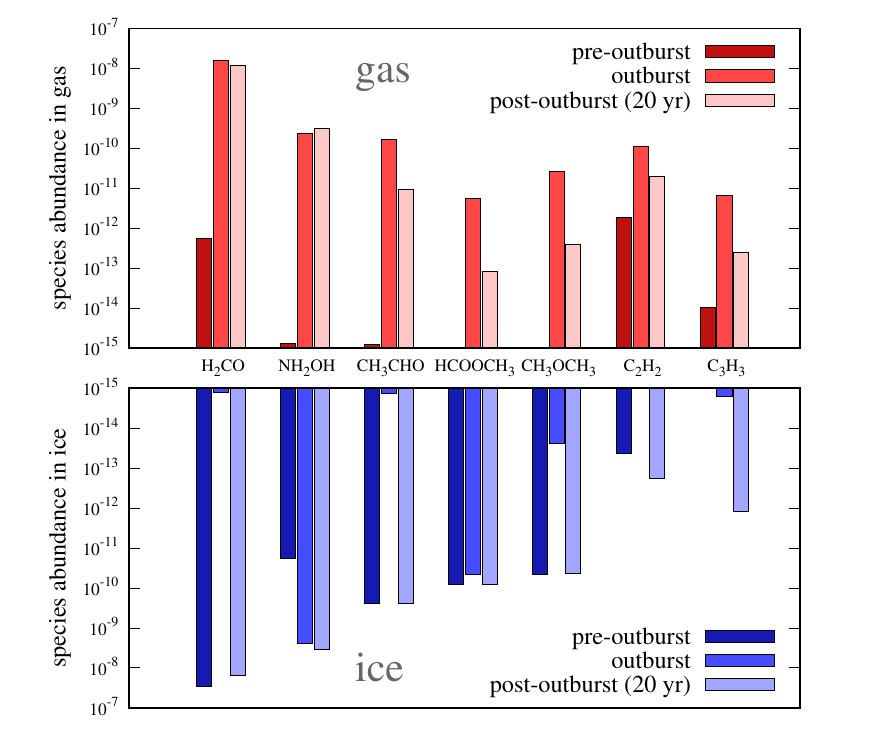}
 \caption{The total disk abundances of assorted outburst tracers in gas and in ice before, during, and 20 years after the outburst (time points 1, 2, and 3 of Figure~\ref{fig:burstprof}, respectively) for the reference disk model. }
  \label{fig:histogram}
\end{figure}

The key chemical processes responsible for the behavior of these selected species are illustrated in Figure~\ref{fig:histogram}, which shows total abundances of gas and ice-phase species in one characteristic disk model. Before the outburst C$_3$H$_3$ and C$_2$H$_2$ are mostly present in the gas; their abundances during the outburst are enhanced through various gas-phase reactions. At the same time, all the other presented species originate from the ices, evaporated by the outburst. After the disk cools down these species, except for hydroxylamine, return to dust surface at the timescales determined by pure freeze-out. NH$_2$OH high abundance in gas is also supported by its efficient formation in ice phase, followed by reactive desorption. These processes are described in more detail in Section~\ref{sec:2d}.

\begin{figure*}
\includegraphics{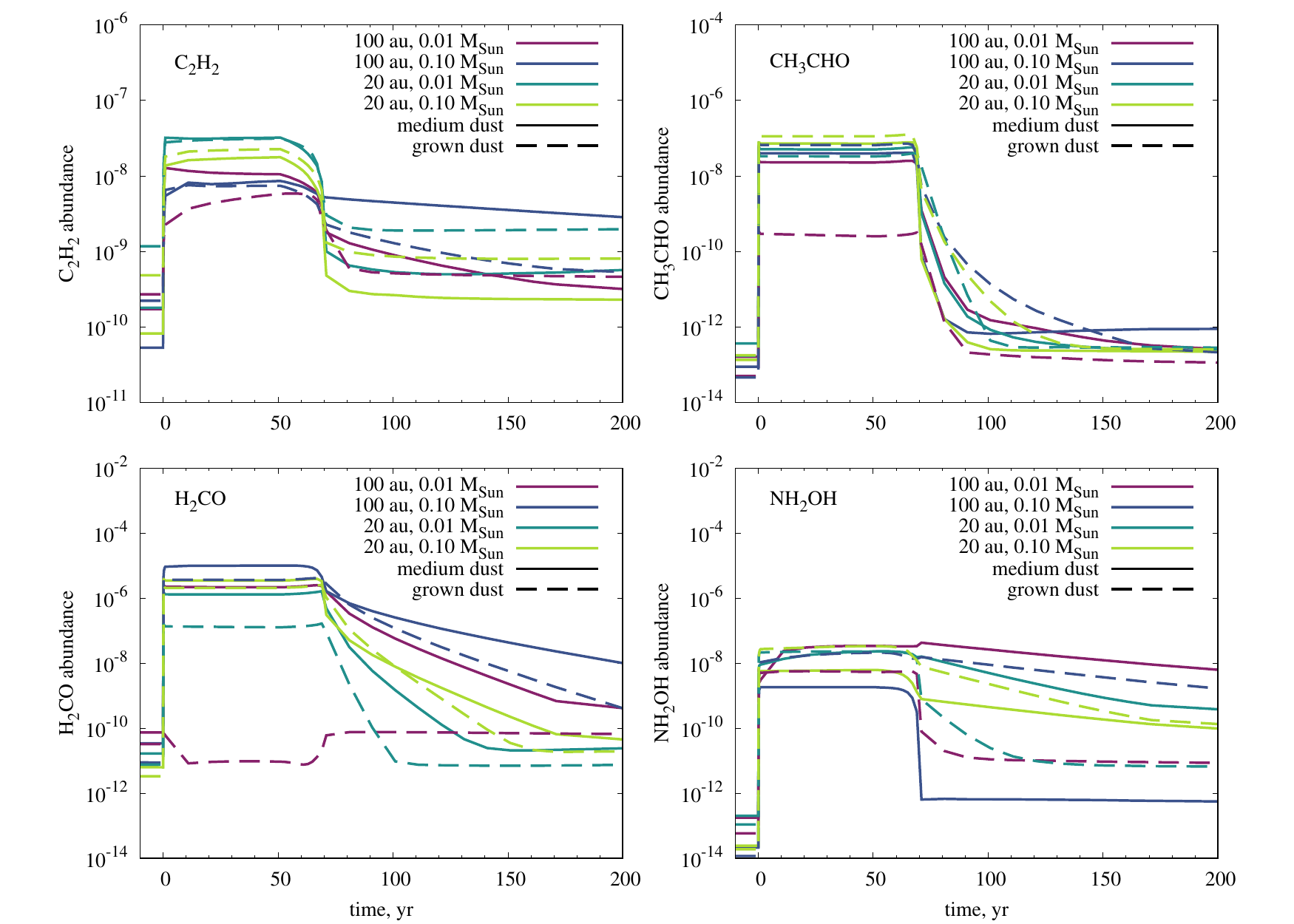}
 \caption{The total disk abundances of assorted outburst tracers during and after the outburst for different disk models. 
 Different lines correspond to the adopted disk mass, characteristic radius, and dust model, as specified in the legend.}
  \label{fig:burstracers}
\end{figure*}

Figure~\ref{fig:burstracers} shows how total disk abundances of selected species change in time in different disk models. All four depicted species behave similarly to CO: their abundances increase by several orders of magnitude at the beginning of the outburst and then fall gradually to the pre-outburst state after its end. We do see, especially in cases of formaldehyde and hydroxylamine, that the decrease of abundance is slow, and in most of the models the molecules are still overabundant at the very end of the outburst, when the luminosity is not enhanced anymore.

We can see differences in models and how each parameter is involved. 
Results for C$_2$H$_2$ are fairly similar between the models, results for CH$_3$CHO vary, but within quite narrow borders, for H$_2$CO and NH$_2$OH the differences are substantial. Note that in models with grown dust (dashed lines in Figure~\ref{fig:burstracers}) the abundances of H$_2$CO and NH$_2$OH drop more rapidly than in  the corresponding models with medium dust (solid lines). For formaldehyde again the model with $M_{\rm disk} = 0.01 M_{\odot}$, $R_{\rm c} = 100$~au, and grown dust (purple dashed line) is noticeable. This is the model corresponding to the arrow in Figure~\ref{fig:timescales}, also it has the highest pre-outburst H$_2$CO abundance. In this model surface formaldehyde is not abundant before the outburst and it is not sublimated from dust during the outburst. Also, photodissociation in the outer disk regions cause its effective destruction during the outburst, which is not compensated by the surface synthesis. As a result, this is one of a few examples of the gas-phase species with the abundance decreasing during the outburst.

Hydroxylamine seems to have two sub-components with two different ways of destruction: one disappears rapidly after the outburst, the other is destroyed extremely slowly. These components are two reservoirs of NH$_2$OH existing under different physical conditions in the disk.
In some models ($M_{\rm disk} = 0.01 M_{\odot}$, $R_{\rm c} = 100$~au and $R_{\rm c} = 20$~au, grown dust; $M_{\rm disk} = 0.1 M_{\odot}$, $R_{\rm c} = 100$~au, medium dust) the first component dominates and we get a group of three models with low $\tau_{\rm ch}$ in Figure~\ref{fig:timescales}; in other models the slowly disappearing component is dominant. This will be farther discussed in the next subsection.

\subsubsection{Two-dimensional chemical structure}
\label{sec:2d}

In this subsection we consider chemical effects of the outburst in more detail and study spatial distribution of selected species.

The differences in species outburst profiles originate from different chemical responses to the outburst. We can distinguish three pathways to the increase of the gas-phase abundance: evaporation from ice phase; formation in the gas phase; formation on ice followed by reactive desorption. The disappearance from the gas-phase may involve a destruction due to photo-dissociation by increased radiation field or chains of reactions that become effective under new physical conditions. 

As it is shown in \citet{W18}, the evaporation of ice mantles is the most important effect for the majority of species. For example, CO, H$_2$CO and CH$_3$CHO abundance changes are explained by pure evaporation/condensation in most of the considered models. The reaction to the outburst in a particular disk in this case depends on initial amount of ices.
The evaporation is very effective and occurs almost instantaneously, while the returning to the surface can be slow.

\begin{figure*}
\includegraphics[width=2\columnwidth]{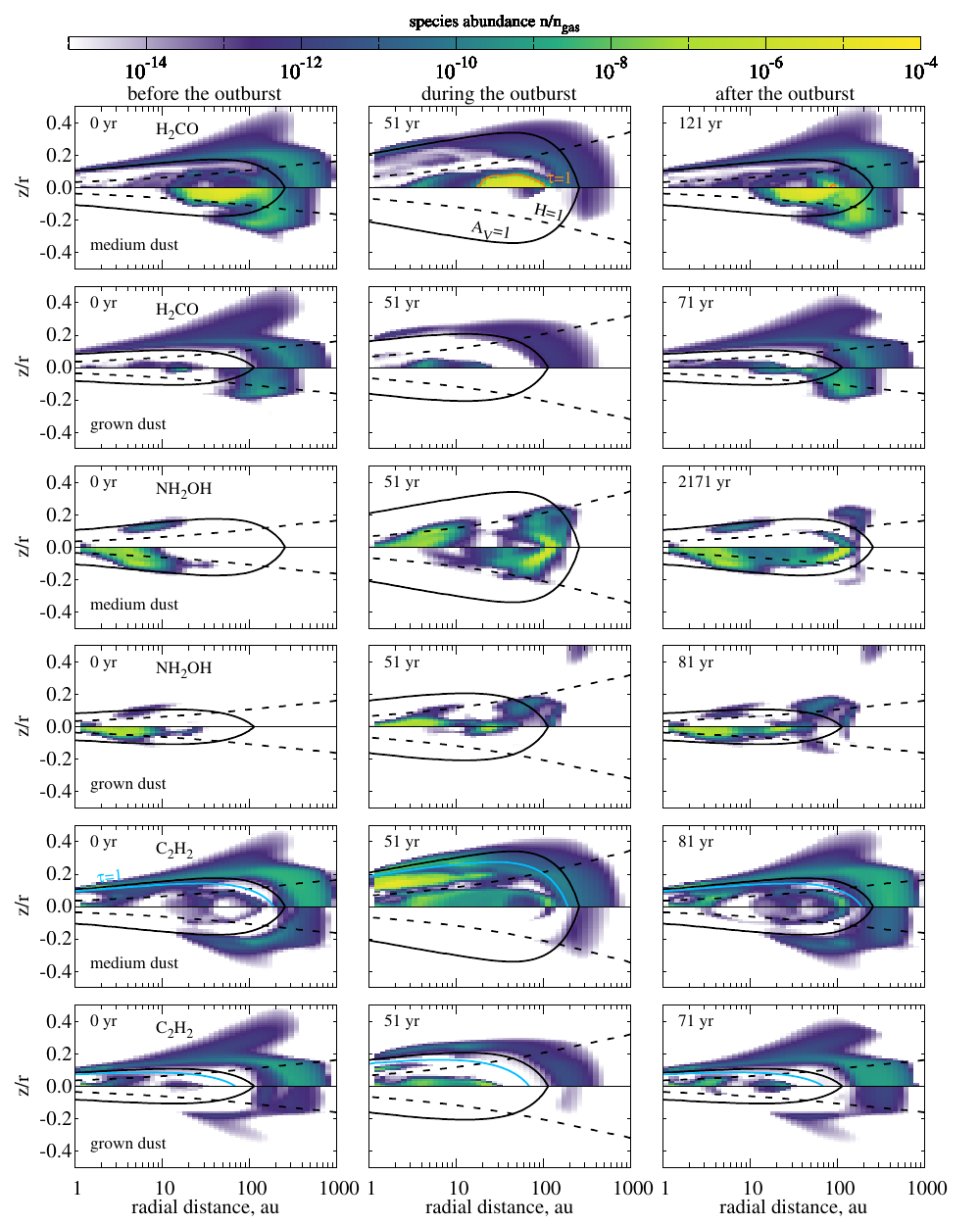}
 \caption{Distribution of selected relative abundances over the disk for two disk models with $R_{\rm c} = 100$~au, $M_{\rm disk} = 0.01 M_{\odot}$, one with medium dust and one with grown dust. Three time points corresponding to moments before (0\,yr), during (51\,yr) and $\tau_{\rm ch}$~years after the outburst are shown. Upper half of each plot depicts the distribution in gas phase, bottom half is for ice phase. The abundances are shown relatively to total gas number density $n_{\rm gas}$. Black solid lines circumscribe the $A_{\rm V}=1$ surface, black dashed lines show the vertical scale height boarder $H=1$, orange lines correspond to $\tau=1$ for $3_{\rm 03}-2_{\rm 02}$ transition of para-H$_2$CO, blue solid lines on C$_2$H$_2$ panels correspond to $\tau_{\rm 3\mu m}=1$. The labeled times are counted from the beginning of the outburst (moment~1 in Figure~\ref{fig:burstprof})}
  \label{fig:2Dtracers}
\end{figure*}

Other species are not simply transferred from solid state to the gaseous one. They form during the outburst either directly in gas, as, e.g., C$_2$H$_2$ and C$_3$H$_3$, or first form on ice and after that pass to gas phase through the reactive desorption or direct evaporation, as NH$_2$OH. Due to ammonia evaporation, hydroxylamine formation on ice starts to dominate, leading to many orders of magnitude increase in hydroxylamine abundance on dust surface. The routes of hydroxylamine formation are described thoroughly in \citet{W18}. 

The examples of spatial distributions for H$_2$CO, NH$_2$OH and C$_2$H$_2$ are shown in Figure~\ref{fig:2Dtracers}. The reference model ($M_{\rm disk}=0.1$\,$M_{\odot}$, $R_{\rm c}=100$\,au) and the one where only the dust size is different are shown. We present abundance distributions for time moments 1 (before the outburst), 2 (during the outburst), and 3 (at the time corresponding to $\tau_{\rm ch}$). Three shown species represent three types of response to the outburst. Formaldehyde shown in the first row is abundant in ice form before the outburst (the left panel), then it is evaporated during the outburst (the middle panel), and after that freezes out again, but even after several decades after the outburst its gas-phase abundance exceeds significantly the pre-outburst level. This is mostly related to slow H$_2$CO freeze-out in low density disk periphery $\gtrsim 100$\,au. In the model with grown dust (the fourth row) formaldehyde ice is nearly absent before the outburst. This is the model which shows a deviant trend in Figure~\ref{fig:burstracers}. 

Hydroxylamine belongs to the second type of species, which form in ice mantle {\it during} the outburst and are reactively desorbed into gas. As shown in the second row in Figure~\ref{fig:2Dtracers}, before the outburst NH$_2$OH ice is only present in the inner disk, at the radial distance of 10~au and less, but during the outburst it forms even at $\sim$100~au. Notably, it does not disappear completely even after 2000~years, being an evidence of irreversible chemical impact of the outburst. Its total abundance still exceeds the pre-outburst value by 2~orders of magnitude even at $10^{4}$\,yr. Additionally, there remain an NH$_2$OH ice reservoir in the inner disk. This ice is evaporated and then rapidly frozen out just like H$_2$CO in medium-dust case. So, the integrated NH$_2$OH  abundance behavior depends on the relation between these two regions. In the case of grown dust (the fifth row) NH$_2$OH ice forms closer to the star, in a warmer region at $\approx$20--30~au, is less abundant and only resides there for a few decades. This explains the two groups of points in Figure~\ref{fig:timescales} for NH$_2$OH and its two-component behavior in Figure~\ref{fig:burstracers}.

The third variety of the outburst impact is represented by C$_2$H$_2$, which is shown in the third and the last rows of Figure~\ref{fig:2Dtracers}. During the outburst it resides in a thick molecular layer in the disk atmosphere as well as in the outer midplane. It forms directly in gas. In the midplane the main formation route is the reaction of C$_3$H$_3$ with O. In the atmosphere C$_2$H$_2$ forms mainly in the reaction of C$_2$H with H$_2$, and via photodissociation of C$_3$H$_2$. The rapid formation process is balanced by photodissociation. C$_3$H$_3$ itself is formed in the reaction of C$_3$H$_4$ with atomic hydrogen, and C$_3$H$_4$ comes to the gas being evaporated by the outburst from the ice phase. In the grown dust model the trends are qualitatively the same. 

Figure~\ref{fig:2Dtracers} shows also optical depth isolines for formaldehyde and acetylene to probe their observability. Orange lines $\tau=1$ are drawn for the formaldehyde transition $3_{\rm 03}-2_{\rm 02}$, which has been detected in several protoplanetary disks (in LkCa\,15 by IRAM 30~m telescope \citep{2004A&A...425..955T}, in AB\,Aur by IRAM/NOEMA \citep{2016A&A...589A..60P}, and in HD\,163296 by ALMA \citep{2017A&A...605A..21C}). During and after the outburst the disk is optically thick in this line at large radial distances. Acetylene, being a symmetric molecule, has no rotational transitions and cannot be observed with ALMA, but it has been detected in disks in its infrared lines around 3\,$\mu$m \citep{2012ApJ...747...92M} and in mid-infrared at around 13\,$\mu$m by {\it Spitzer} and TEXES \citep{2008Sci...319.1504C,2011ApJ...733..102C,2018arXiv180703406N}. Observability in IR can be assessed by dust continuum optical depth: the molecular emission would be observable only if it comes from above $\tau_{\rm 3\mu m}=1$ surface.

\subsubsection{Peculiar species}
\label{sec:methanolnco}

The adopted way to calculate retention timescale $\tau_{\rm ch}$ is not suitable for species with non-monotonic behavior of post-outburst abundances. Among them are CH$_3$OH, CH$_5$N and HNC molecules. Their abundances follow the luminosity drop and return very rapidly to low values, but after that start to grow again, which results in them being up to an order of magnitude overabundant after hundreds of years. Hydroxyl is also concerned as peculiar, as the gas-phase species with abundance notably falling immediately after the outburst end.

\begin{figure*}
\includegraphics[width=2.1\columnwidth]{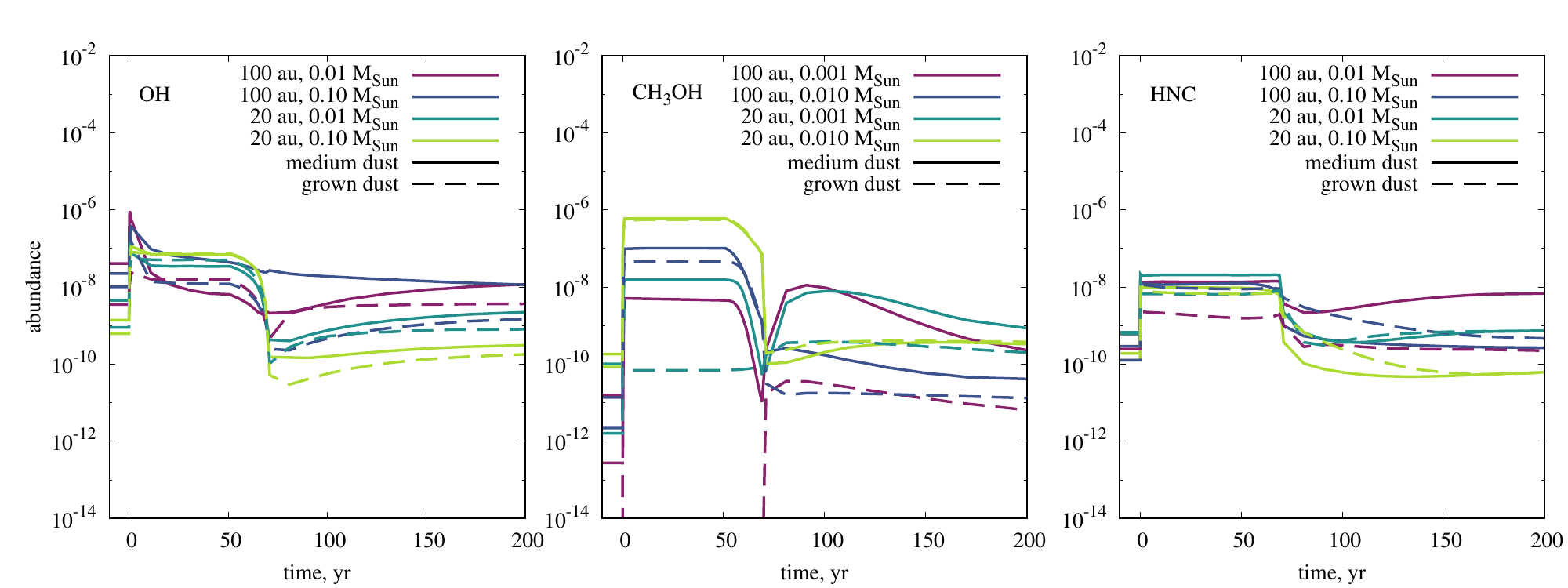}
 \caption{Average abundances of peculiar species during and after the outburst for different models. Unlike everywhere else in this work, CH$_3$OH is shown for models with $M_{\rm disk} = 0.01$ and $0.001 M_{\odot}$, as its behavior is especially interesting in the models with low-mass disks.}
  \label{fig:pectracers}
\end{figure*}

Figure~\ref{fig:pectracers} shows the abundances of hydroxyl, methanol, and hydrogen isocyanide, changing during the outburst. Methanol behavior is especially interesting in the models with low-mass disks, so disk masses of $0.01 M_{\odot}$ and $0.001 M_{\odot}$ are plotted. In these models CH$_3$OH abundance after the outburst exceeds the pre-outburst abundance by~2--3~orders of magnitude, and may also exceed the outburst abundances, especially in grown dust models.

\section{Discussion}

To check the observability of H$_2$CO before, during and after the outburst, we performed line radiative transfer (LRT) modeling using one of the brighest formaldehyde lines in disks, the para-H$_2$CO ($3_{03}-2_{02}$) transition at 218.222\,GHz. This transition can be detected in disks even when the formaldehyde abundances are low, $\sim 10^{-10}$ \citep[wrt to H;][]{2017A&A...605A..21C}; low upper state energy of $10.956$\,K allows it to emit even in very cold disk regions. We used the calculated H$_2$CO abundances in the disk at $t=0$, $51$, and $121$~years after the beginning of the outburst and assumed an equilibrium ortho-to-para ratio of 3:1. The 3D LTE LRT calculations were done using the LIne Modelling Engine (\textsc{lime}) \citep{Brinch2010} v.~1.3 and the para-H$_2$CO level population data from the Leiden Atomic and Molecular Database \citep[LAMDA,][]{Schoier2005}.

In order to make our modeling as realistic as possible, some input parameters were based on the FUor-type outburst occurred in the V346~Nor system. V346~Nor is a young embedded protostar at a distance of 700\,pc. It brightened significantly around 1980, which was soon classified as a FUor-type outburst, and remained in the high state for decades \citep{2017A&A...597L..10K}. In 2010, however, the outburst suddenly ended, making V346~Nor the first and only known FUor that has returned to quiescence up to now \citep{2016MNRAS.462L..61K}.  It should be noted that while our model includes no envelope, V346~Nor is an embedded object, showing 10\,$\mu$m silicate feature in absorption \citep{2007ApJ...668..359Q}, so it is not well-modeled by a bare disk. Although objects with tenuous envelopes, like FU~Ori itself \citep{2015ApJ...812..134H} or HBC~722 \citep{2011ApJ...731L..25G} could be better for the comparison, we choose V346~Nor, as it presents a unique example of a post-outburst FUor.
For our final modeling we adopted a distance of 700\,pc, as the disk inclination is not known, we choose $30\degr$ assuming random disk orientation. The time evolution of the outburst was approximated by the relationship in Figure~\ref{fig:burstprof}. The used stellar mass was $0.4$\,$M_{\odot}$, slightly higher than estimated for V346~Nor from recent ALMA observations \citep{2017ApJ...843...45K}, but still representative of the class of eruptive protostars. For the disk parameters we adopted those of our reference disk model (Figure~\ref{fig:2Dtracers}).

\begin{figure}
\includegraphics[width=0.45\textwidth]{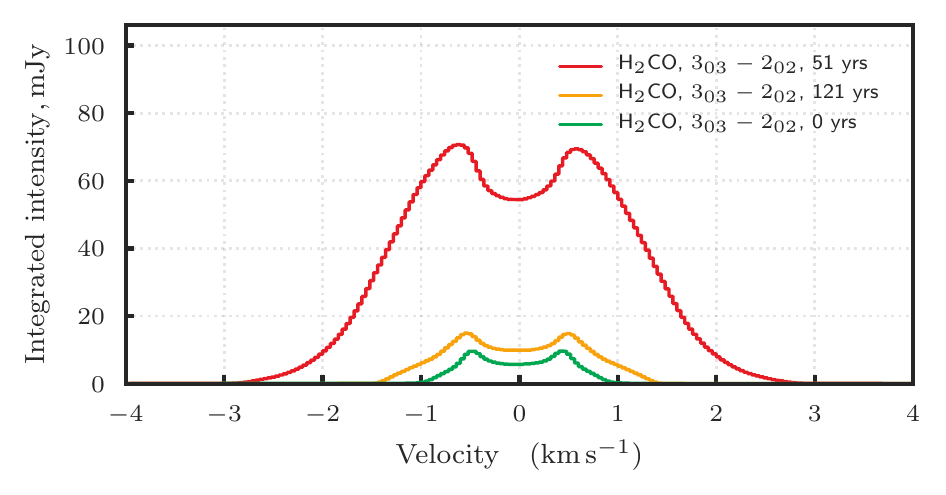}
 \caption{Integrated line intensities of H$_2$CO $3_{\rm 03}-2_{\rm 02}$ transition for the reference model at time points shown in Figure~\ref{fig:2Dtracers}, the top row.}
  \label{fig:lines}
\end{figure}

The calculated integrated intensity of this line at the three evolutionary moments are shown in Figure~\ref{fig:lines}. As can be clearly seen, the line intensity is the highest, $\sim 60$~mJy\,km\,s$^{-1}$, after about several tens of years since the onset of the outburst. Using ALMA Cycle~6 OT Sensitivity Calculator, we estimated that this line can become detectable by ALMA at a spectral resolution of $0.3$\,km\,s$^{-1}$ after a few hours of the on-source integration; according to NOEMA/PolyFix exposure calculator the detection by NOEMA is possible with several hours of integration.

The molecular phase preceding the quiescent disk period is very important because during this time most of organic ice mantles are formed. The duration of this stage as well as pre-outburst evolution may affect the composition of the material evaporated during the outburst. The dust size also changes discontinuously between cloud and disk stages, which may lead to non-equilibrium ice composition which is unimportant until this ice is released into gas phase by sudden temperature rise. Having less surface of the grown dust ensemble, all the species cannot return to ice, and enhance the applicability of listed tracers.

This effect may be compensated by multiple outbursts, which give the chemical equilibrium another chance to restore. Also serial outbursts may serve as a molecular factory, allowing complex organic molecules be produced with higher efficiency. This effect is to be examined thoroughly in future studies.

One of the species pointed out by \citet{W18} is benzene (C$_6$H$_6$), which is effectively formed in ice after the outburst in the inner and intermediate disk regions. In this work, we do not see benzene as one of the outburst tracers, because the region where it forms is small. It does form in the midplane at $\approx$10~au, but this increase is not noticeable for the whole disk.

Ideally, we would like to determine dust properties in the disk by observations of some molecules, whose abundances are sensitive to dust size. However, despite that different dust sizes do lead to different chemical outcomes of the outburst, the variations caused by the disk physical structure are large, too. The differences between models due to disk size and mass are comparable with those caused by changed dust size, so there is no distinct chemical tracer of dust size in the present model, both for immediate and long-term tracers. For the former this can be seen from Table~\ref{tab:burstmarkers}, as total disk abundances for medium and grown dust models are quite similar. For the long-term tracers, we would expect that for a given gas temperature and density larger average grain size, i.e., less total grain surface area and lower recondensation rates, leads to species more slowly returning to the pre-outburst abundances. It is true in some cases (see, e.g., CH$_3$CHO and NH$_2$OH in models with $M_{\rm disk}=0.1$\,$M_{\odot}$, Figure~\ref{fig:burstracers}), however, for some species the trend is opposite, e.g., for C$_2$H$_2$ in the model with $R_{\rm c}=$100\,au, $M_{\rm disk}=0.1$\,$M_{\odot}$, and for H$_2$CO in all models. This diversity arises from the fact that different dust properties affect not only available grain surface, but also disk temperature and radiation field. This may be further blurred by the variety of reactions contributing to the considered species abundance. We conclude that there is no no well-defined trend governed by dust size, at least for the selected molecules.

Noteworthy is that the obtained chemical effects are to some degree model-dependent. Updates of astrochemical networks with new laboratory data, as well as more sophisticated models, may result in different molecular abundances, especially for complex species. The long-term molecular line observations of erupting young stars thus provide a nice opportunity to calibrate astrochemical models.

The observability of selected tracers needs to be examined in radiation transfer modeling. Plenty of species that are currently detected in disks have very low abundances (see Table~\ref{tab:burstmarkers}, bottom). This indicates that high abundances are not necessary, and the species can still be bright if the excitation conditions are satisfied. Thus, many species that could actually be detected might be rejected by our criteria. 

\section{Conclusions}

We carried out the modeling of protoplanetary disk chemical composition affected by the FUor-like luminosity outburst. Our goal was to find chemical signatures of past outburst left in the disk after the end of the outburst. Our results can be summarized as follows:

1. CO abundance does not retain the outburst influence for a long time after the outburst in the absence of an embedding envelope. Variations of CO abundance among disks with different physical parameters are comparable to those caused by the outburst. In combination with outburst tracing species CO can be used as a reference molecule to identify the species abundance change.

2. We find species which are particularly sensitive to the outburst with abundances growing by two and more orders of magnitude. These species are listed in Table~\ref{tab:burstmarkers}, among them the most noticeable are CH$_3$OCH$_3$, CH$_3$CHO, NH$_2$OH, and HC$_5$N, which have never been observed in ``normal'' disks, but can reach high abundances in case of the outburst.

3. Some species can be used to identify past outbursts in quiescent disks. H$_2$CO is the most promising species for recently ended outbursts (30-120\,yr). NH$_2$OH abundance remains more than three order of magnitude above its pre-outburst value for thousands of years after the outburst. In disks with grown dust the timescale is several decades. 

4. There are three main mechanisms responsible for the total abundance evolution. The first one is evaporation of ices followed by freeze-out, as for CO, H$_2$CO, C$_2$H$_6$ etc. The second one is formation on the dust surface during the outburst, as for NH$_2$OH. The third is the formation in the gas-phase, as for C$_2$H$_2$ and C$_3$H$_3$.

5. Dust size can influence the chemical outcome of the outburst, but in an ambiguous way, leaving no distinct features capable of indicating it by chemical changes in the disk.

\section*{Acknowledgements}
TM, VA, EV and DW acknowledge support from the RFBR grant 17-02-00644. PA and AK acknowledge funding from the European Research Council (ERC) under the European Union`s Horizon 2020 research and innovation programme under grant agreement No 716155 (SACCRED).

\software{ANDES \citep{Akimkin13}}

\bibliographystyle{aasjournal}
\bibliography{refs}

\end{document}